%Paper: 9112053
%From: <XIONG%ITSSISSA.BITNET@ICINECA.CINECA.IT>
%Date: Thu, 19 Dec 91 16:24 +0100

%%%%%%%%%%%%%%%%%%%%%%%%%%%%%%%%%%%%%%%%%%%%%%%%%
%         This is phyzzx file                   %
%%%%%%%%%%%%%%%%%%%%%%%%%%%%%%%%%%%%%%%%%%%%%%%%%
\input phyzzx.tex
\titlepage
\hfill{SISSA-ISAS 187/91/EP}
\title{Generalized Integrable Lattice Systems}
\author{ C.S.Xiong}
\address{ International School for Advanced Studies (SISSA/ISAS),
Strada Costiera 11, I-34014 Trieste, Italy}
\vskip .6truein
\abstract
We generalize Toda--like integrable lattice systems to
non--symmetric case. We show that they possess
the bi--Hamiltonian structure.

\vskip .4truein

\endpage
\pagenumber=1

\section{Introduction}

In the last two years there has been a great progress in understanding
two dimensional quantum gravity coupled to mimimal conformal models.
One of the powerful tool to attack this problem is the so--called matrix
model. It is well--known that one--matrix model is deeply connected
with Toda--like (or Volterra--like ) Lattice system[1,2,3], which is the
simplest classical integrable model on the lattice[4,5].
Furthermore, on it one can realize a discretized Virasoro algebra
as well as discretized KdV--equations, and probably it is related to
a lattice version of Liouville theory[3,6].

These properties make it a natural question to consider generalizations
of this model, maintaining the integrability, in particular, playing roles
in the multi--matrix model and giving the possible formulation of the
discrete $W_n$--algebra. A generalization has been proposed
in[7], it is still an integrable symmetric lattice, but unfortunately it
seems to have nothing to do with multi--matrix models. In this letter,
we analyze a non--symmetric lattice and its associated discrete linear
system. We prove that it is integrable. Its important role
in multi--matrix models will be presented elsewhere[8].

\section{Notations}

Let us present here our notations.
For any  matrix M, we denotes its upper and lower triangular
parts by $M_{\pm}$ respectively, and by $M_{(0)}$ means the diagonal line.
We also introduce the definitions
$$
M_{\theta}\equiv {1\over2}M_{(0)}+M_-,\qquad\qquad
M_{\alpha}\equiv {1\over2}M_{(0)}+M_+
$$$$
M_a\equiv M_{(0)}+M_+,
\qquad\qquad
I_{\pm}\equiv \sum_{i=0}^{\infty}E_{i,i\pm 1},
$$
As usual $(E_{ij})_{kl}=\delta_{ik}\delta_{jl}$. We will consider
throughout semi--infinte
matrices, $0\geq i,j<\infty$. Furthermore, for the later convenience,
we introduce the natural gradation
$$
deg[E_{ij}]\equiv j-i
$$
$M_{[a,b]}$ denotes the diagonal lines of matrix M
with degrees from ``a" to ``b", while $M\in[a, b]$ means that
the degrees of matrix M is only from ``a" to ``b".

As we know, under matrix commutation, the semi--infinite matrices
form an infinite dimensional algebra, we denote it by $gl_0(\infty)$,
which has the following decomposition
$$
gl_0(\infty)={\cal N}_+\oplus{\cal H}\oplus{\cal N}_-
$$
the subalgebra ${\cal N}_+$(${\cal N}_-$) contains all the upper(lower)
triangular matrices, while ${\cal H}$ represents the Abelian subalgebra
formed by diagonal matrices. Furthermore, we introduce two
subalgebras of $gl(\infty)$
$$
{\cal P}_-\equiv {\cal H}\oplus{\cal N}_-, \qquad\qquad
{\cal P}_+\equiv {\cal H}\oplus{\cal N}_+
$$

\section{The Symmetries of The Discrete Linear Systems}

Let us begin with the following discrete linear system
$$
\cases{
Q\xi=\lambda\xi\cr
{\partial\over{\partial t_r}}\xi=(Q^r)_{\theta}\xi\cr}
\eqno\hbox{(1)}
$$
where the Jacobi matrix takes the following explicit form
$$
Q_{ij}=\sqrt{R_j}E_{i,i+1}\delta_{j,i+1}
+\sum_{l\geq 0}^n L^{(l)}_i E_{i,i-l}\delta_{j,i-l},\qquad\qquad i,j\geq 0
\eqno\hbox{(2)}
$$
Evidantly $Q\in[-n,1]$ with arbitary positive integer $n$,
while $R_i$'s and $L^{(l)}_i$'s are coordinate variables of the
system, whose equations of motion are
$$
{\partial\over{\partial t_r}}Q=[(Q^r)_{\theta},~~Q]\qquad\qquad
\forall r\geq 1
\eqno\hbox{(3)}
$$
They are just the consistency conditions of the system (1), and
we will call them the discrete KdV equations.

In this section we will consider two different types of the symmetries
of the generalized discrete linear system: a gauge symmetry and a Weyl
symmetry. The gauge
symmetry preserves completely the dynamics of the system, while
the Weyl scaling can be used to reduce the system (1) to the
standard form.

\subsection{Gauge Symmetry of The System}

By gauge symmetry we mean the invariance of the system under the
following transformations
$$
\cases{Q\longrightarrow {\tilde Q}=G^{-1}QG\cr
\xi\longrightarrow {\tilde \xi}=G^{-1}\xi\cr}
\eqno\hbox{(4)}
$$
where G is some invertible matrix such that the form of the system~(1)
remains unchanged, i.e.
$$
\cases{
{\tilde Q}{\tilde \xi}=\lambda{\tilde \xi}\cr
{\partial\over{\partial t_r}}{\tilde \xi}
=({\tilde Q}^r)_{\theta}{\tilde \xi}\cr}
$$
That is to say, the dynamics of the system does not
change. In order to see what this symmetry is,
we consider its infinitesimal form
$G=1+\epsilon g$.
The invariance requires that the matrix g should satisfy the equations
$$
\cases{
{\tilde Q}=Q+\epsilon[Q, g],\cr
{\partial\over{\partial t_r}}g=[Q^r_{\theta}, g]-[Q^r, g]_{\theta}\cr}
\eqno\hbox{(5)}
$$
Two solutions are
$$
g_1=\sum_{k}b_kQ^k, \qquad g_2=\sum_{k}c_kQ^k_{\alpha},
\qquad \forall k\geq 1
\eqno\hbox{(6)}
$$
where $b_k$'s and $c_k$'s are time--independent constants.
The effect of the transformations generated by $g_1$'s is only to
rescale the basis by a $\lambda$--dependent factor, while the
transformations generated by the $g_2$'s is in fact equivalent to
tune the time parameters $t_r$'s. They
form  an infinite dimensional gauge group.
We will see that,
it is just this symmetry that leads to the integrability
of the discrete linear system.

\subsection{Weyl Scaling Symmetry}

Another kind of symmetry of the system (1) is Weyl scaling
invariance. Suppose that we choose a diagonal matrix G.
After the transformation (4), the time evolution equations of $\xi$
will transform to another form, but the equations of motion of the
coordinates $R_i$'s and $L^{(l)}_i$'s keep unchanged. The only
effect of this transformation is to rescale the basis of the vector
$\xi$ by a $\lambda$--independent factor.
This symmetry has only one degree of freedom. Now let us
choose the following special gauge
$G_{ij}={1\over {\sqrt{h_j}}}\delta_{ij}$( where we have introduced
the quantities $h_j/h_{j-1}=R_j$, $h_0$ is an undetermined constant which
is not important in our analysis), then the Jacobi matrix becomes
$$
Q=I_++\sum_{l=0}^n A^{(l)},\qquad\qquad
A^{(l)}\equiv\sum_{i=l}^{\infty} A^{(l)}_iE_{i,i-l}
\eqno\hbox{(7a)}
$$
with
$$
A^{(0)}_j=L^{(0)}_j,\qquad A^{(l)}_j=L^{(l)}_j\sqrt{R_jR_{j-1}\ldots
R_{j-l+1}},\qquad \forall l\geq 1
\eqno\hbox{(7b)}
$$
In this gauge, it can be easily checked that the linear system (1)
takes the following form
$$
\cases{
Q\xi=\lambda\xi\cr
{\partial\over{\partial t_r}}\xi=(Q^r)_-\xi\cr}
\eqno\hbox{(8)}
$$
The equations of motion (or consistency conditions of the system (8))
read
$$
{\partial\over{\partial t_r}}Q=[(Q^r)_-,~~Q]\qquad\qquad
\forall r\geq 1
\eqno\hbox{(9)}
$$
We see that the equations of motion of the coordinates $R_j$'s
disappear, in fact it is implicitly involved in the equations
of motion of the coordinates $A_j$'s.
Both system (1) and system (8) appear in multi--matrix models[8].

\section{Integrability of The Discrete Linear System (8)}

As we know a dynamical system with $n$ degrees of freedom is
integrable if and only if there are $n$--independent conserved
quantities in involution[4,5]. For a system with infinite many
degrees
of freedom, there should be infinite many independent conserved
quantities in involution. One of the main approaches to show the
integrability is the so--called bi--Hamitonian method, i.e.
one should prove that there exist at least two compatible Poisson
brackets. Compatibility of two Poisson brackets means
$$
\{ H_{k+2}, f_Y\}_1(Q)=\{H_{k+1}, f_Y\}_2(Q)
\eqno\hbox{(10)}
$$
for any Hamiltonian $H_k$. Hereafter we will follow this line to
construct two compatible Poisson brackets, which gives the discrete
KdV equations (9).

\subsection{The First Poisson Bracket}

Our strategy is as follows. We try to define such kind of Poisson
structure, which gives the desired equations of motion, i.e.
the discrete KdV--equations (9). In order to do this, let us
define the trace operation on the matrix space
\footnote{*}{we suppose that this infinite summation is well--defined}
$$
Tr(M)\equiv\sum_{i=0}^{\infty}M_{ii}
$$
which gives a natural inner scalar product on $gl_0(\infty)$.
It is easy to see that the product is symmetric and invariant
under the action of $gl_0(\infty)$
$$
\cases{
Tr(AB)=Tr(BA)\equiv A(B)=B(A)\cr
Tr(A[B, C])=Tr([A, B]C)\cr}
$$
with respect to this product, ${\cal P}_+$ and ${\cal P}_-$ are
dual algebras, their elements being in 1--1 correspondence. On the
other hand, the functions of $A^{(l)}_j$'s, which span a functional
space ${\cal F}$, can be defined as
$$
f_X(Q)\equiv Tr(QX), \qquad\qquad df_X(Q)=X\in {\cal P}_+
$$
This is a function on the algebra ${\cal P}_-$. Consequently
 the algebraic
structure of ${\cal P}_-$ determines the Poisson structure on
${\cal F}$ through the Konstant--Kirillov  bracket
$$
\{f_X, f_Y\}_1(Q)=Q([X, Y]),\qquad\qquad X,Y\in{\cal P}_+
\eqno\hbox{(11)}
$$
Of course it is anti--symmetric and satisfies the Jacobi identity.
Furtheremore, one can show that with respect to this Poisson
bracket, the conserved functions are
$$
H_k={1\over k}Tr(Q^k)\qquad
dH_k=Q^{(k-1)}_{[0,n]}
\qquad \forall k\geq 1
\eqno\hbox{(12)}
$$
They are in involution, i.e.
$$
\{H_{k+1}, H_{l+1}\}_1(Q)=0
$$
So they can be chosen as Hamiltonians, and give the desired equations
of motion, i.e. the discrete KdV--Heirachies. On the other hand, these
Hamiltonians generate the symmetry transformations on the
functional space ${\cal F}$. In order to see its relation to the
gauge transformations (6),
we recall that the co--adjoint action of ${\cal P}_+$ on the
functional space ${\cal F}$ is defined through its adjoint action on
the coordinates space ${\cal P}_-$, i.e.
$$
ad^{*}_Y f_X(Q)\equiv f_X(ad_Y(Q))=X([Y, Q])\qquad X,Y\in{\cal P}_+
$$
Obviously the symmetry on ${\cal F}$ generated by the Hamiltonians
is nothing but the gauge symmetry on the coordinate space.
Therefore, for a system with infinite degrees of freedom,
there must exist an infinite dimensional gauge symmetry, such
that we could find infinite many independent involutive conserved
quantities.

\subsection{The Second Poisson Bracket}

In order to get the second Poisson bracket, our starting point is
compatibility of two Poisson brackets.
If we can represent the right hand side of eq.(10)
 in terms of $dH_{k+1}$, then
we can extract the second Poisson bracket immediately.
Consider the following matrix
$$
\eqalign{
F&=[Q, Q^{k+1}_a]\cr
&=\bigl(Q(Q^k_aQ)_a-(QQ^k_a)_aQ\bigl)
+\bigl(QQ^k_{[-1]}I_+-I_+Q^k_{[-1]}Q\bigl)\cr}
$$
{}From the equality
$$
[Q, Q^k_a]=[Q^k_-, Q]
$$
we obtain
$$
Q^k_{[-1]}
=\sum_{j=0}^{\infty}Q^k_{j+1,j}E_{j+1,j},\quad
Q^k_{j+1,j}
=\sum_{l=0}^j[dH_{k+1}, Q]_{ll}
$$
Therefore we get
$$
\eqalign{
\{H_{k+2}, f_Y\}_1(Q)&=Q([dH_{k+2}, Y])=Y(F)\cr
&=<(dH_{k+1}Q)_aYQ>-<(QdH_{k+1})_aQY>\cr
&+\sum_{i,j}Q_{ij}(Q^k_{j,j-1}-Q^k_{i+1,i})Y_{ji}\cr}
$$
substituting $dH_{k+1}$ by X, and introduce a X--dependent matrix
$$
{\cal D}\equiv \sum_{j=0}^{\infty}{\cal D}_jE_{j+1,j}\qquad\qquad
{\cal D}_j\equiv \sum_{l=0}^j[X, Q]_{ll}
\eqno\hbox{(13)}
$$
we finally obtain the second Poisson bracket
$$
\eqalign{
\{f_X, f_Y\}_2(Q)&=<(XQ)_aYQ>-<(QX)_aQY>\cr
&+<Q{\cal D}I_+Y>-<QYI_+{\cal D}>\cr}
\eqno\hbox{(14)}
$$
Writing explicitly in terms of the coordinates $A^{(l)}_j$'s, we have
$$
\eqalign{
\{ A^{(a)}_i, A^{(b)}_j\}_2&=
A^{(a+b+1)}_j\delta_{i,j-b-1}-A^{(a+b+1)}_i\delta_{i,j+a+1}\cr
&+\sum_{l=0}^a\bigl(A^{(l)}_iA^{(a+b-l)}_j\delta_{i,j-b+l}
-A^{(a+b-l)}_iA^{(l)}_j\delta_{i,j+a-l}\bigl)\cr
&+ A^{(a)}_iA^{(b)}_j\sum_{l=j-b}^j(\delta_{il}-\delta_{i-a,l})\cr}
\eqno\hbox{(15)}
$$
here we understand $A^{(r)}_i=0$, if $r>n$.
It is tiresome but straightforward  to prove that the above
Poisson bracket is antisymmetric and satisfies the Jacobi identity.
The following is a simple example for $n=2$ and
$$
Q_{ij}=\delta_{j,i+1}+S_j\delta_{ij}+
L_j\delta_{j,i-1}+A_j\delta_{j,i-2}
$$
Eq.(15) gives us the following
Poisson brackets
$$
\eqalignno{
\{S_i, S_j\}_2&=L_j\delta_{i,j-1}-L_i\delta_{i,j+1},&(16a)\cr
\{S_i, L_j\}_2&=S_iL_j(\delta_{i,j-1}-\delta_{ij})
+A_j\delta_{i,j-2}-A_i\delta_{i,j+1},&(16b)\cr
\{S_i, A_j\}_2&=S_iA_j(\delta_{i,j-2}-\delta_{ij}),&(16c)\cr
\{L_i, S_j\}_2&=L_iS_j(\delta_{ij}-\delta_{i,j+1})
+A_j\delta_{i,j-1}-A_i\delta_{i,j+2},&(16d)\cr
\{L_i, L_j\}_2&=L_iL_j(\delta_{i,j-1}-\delta_{i,j+1})
+S_iA_j\delta_{i,j-1}-A_iS_j\delta_{i,j+1},&(16e)\cr
\{L_i, A_j\}_2&=L_iA_j(\delta_{i,j-2}+\delta_{i,j-1}
-\delta_{ij}-\delta_{i,j+1}),&(16f)\cr
\{A_i, S_j\}_2&=A_iS_j(\delta_{ij}-\delta_{i,j+2}),&(16g)\cr
\{A_i, L_j\}_2&=A_iL_j(\delta_{i,j-1}+\delta_{ij}
-\delta_{i,j+1}-\delta_{i,j+2}),&(16h)\cr
\{A_i, A_j\}_2&=A_iA_j(\delta_{i,j-2}+\delta_{i,j-1}
-\delta_{i,j+1}-\delta_{i,j-2}),&(16i)\cr}
$$
{}From our construction of the Poisson brackets,
at the first glance it seems to be possible to deduce higher
order Poisson brackets. However it is very difficult to work it out,
and, furthermore, the deduced higher order Poisson brackets probably
do not satisfy Jacobi identity. The only exception is the ansatz
$n=1$, $r=2$, in this case, we denote
$$
Q=I_++S+R
$$
Then playing  the same game as before,
we can get the third Poisson bracket
of the Toda--like lattice
$$
\eqalign{
\{f_X, f_Y\}_3(Q)&={1\over2}\bigl(<(XQ^2)_aYQ>-
<(Q^2X)_aQY>\cr
&+<(QXQ)_aYQ>-<(QXQ)_aQY>\cr
&<YQC_1>-<QYC_2>\bigl)\cr}
\eqno\hbox{(17)}
$$
where
$$
\eqalign{
B&\equiv \sum_{j=0}^{\infty}B_jE_{j+2,j}, \qquad\qquad
B_j\equiv\sum_{l=0}^j[X, Q_a]_{l+1,l}\cr
{\cal D}&\equiv \sum_{j=0}^{\infty}{\cal D}_jE_{j+1,j},\qquad\qquad
{\cal D}_j\equiv \sum_{l=0}^j[X, Q]_{ll}\cr}
$$
and
$$
\eqalign{
C_1&\equiv BI_+^2+I_+BI_++{\cal D}Q^2_+
2I_+{\cal D}I_+
+I_+{\cal D}S+S{\cal D}I_+\cr
C_2&\equiv I_+^2B+I_+BI_++Q^2_+{\cal D}
+2I_+{\cal D}I_++I_+{\cal D}S+S{\cal D}I_+\cr}
$$
The Poisson brackets of the coordinates are as follows
$$
\eqalign{\{ R_i, R_j\}_1=&0,\qquad\{ S_i, S_j\}_2=0.\cr
\{ R_i, S_j\}_1=&R_i(\delta_{j,i+1}-\delta_{i,j}) \cr}
\eqno\hbox{(18)}
$$
and
$$
\eqalignno{\{ R_i, R_j\}_2 =&R_iR_j (\delta_{j,i+1}-\delta_{i,j+1})
& (19a)\cr
\{ R_i, S_j\}_2=&R_iS_j(\delta_{j,i+1}-\delta_{i,j}) & (19b)
\cr
\{ S_i, S_j\}_2=&R_i\delta_{j,i+1}-R_j\delta_{i,j+1}.
& (19c)\cr}
$$
as well as
$$
\eqalignno{\{R_i,R_j\}_3 =& 2 R_iR_j ( S_i \delta _{i,j-1}- S_j \delta_{i,j+1})
&(20a)\cr
\{R_i , S_j\}_3 =& R_i R_j(\delta _{i,j-1} + \delta _{i,j}) -R_iR_{j+1}
( \delta _{i,j+1}+ \delta _{i,j+2}) \cr
&+ R_i S_j^2(\delta_{i,j}- \delta_{i,j+1}) &(20b)\cr
\{S_i,S_j\}_3 =& (S_i+S_j) (R_j \delta_{i,j-1}- R_i \delta_{i,j+1})&(20c)\cr}
$$
These Poisson brackets have been derived in [3] by a straightforward
calculation, while now they have been obtained
from the sytematic analysis.
In the continuum limit, if we set the infinitesimal parameter
$\epsilon={1\over N}$, and suppose that
$$
\cases{
R_j\longrightarrow 1+{1\over2}\epsilon^2(u(x)+v(x))\cr
S_j\longrightarrow 2+{1\over2}\epsilon^2(u(x)-v(x))\cr}
\eqno\hbox{(3.27)}
$$
after introducing a new Poisson bracket
$$
\{, \}\equiv{1\over4}\{, \}_3-\{, \}_2
$$
we get two copies of Virasoro algebras
$$
\eqalignno{
\{u(x), u(y)\}&=
{1\over2}(\partial^3+4u(x)\partial+2u^{'}(x))\delta(x-y)
&(3.28a)\cr\noalign{\vskip12pt}
\{v(x), v(y)\}&=-
{1\over2}(\partial^3+4v(x)\partial+2v^{'}(x))\delta(x-y)
&(3.28b)\cr\noalign{\vskip12pt}
\{u(x), v(y)\}&=0.&(3.28c)\cr}
$$

\section{Integrability of The System (1)}

In the previous section,  we have gauged away the Weyl symmetry
keeling in this way one degree of freedom. So the coordinates $R_j$'s
do not appear in the Poisson brackets as well as in the discrete KdV
equations. Now let us see what is the suitable Poisson brackets
including them explicitly.

If we do not fix the Weyl symmetry, then we have the discrete
linear system (1), whose integrability can be analysed
in almost the same way  as before. The only difference is that in this
case $Q\in [-n,1]$ but the  tangent vector belongs to
$[-1,n]$, so we should make use of R--matrix of $gl_0(\infty)$,
which acts in the following way
$$\cases{{\cal R}(E_{ij})=E_{ij},\quad\qquad\qquad i<j\cr
{\cal R}(E_{ii})=0\cr
{\cal R}(E_{ij})=-E_{ij},\qquad\qquad i>j.}
\eqno\hbox{(21)}
$$
which defines the following Lie--algebraic structure
on the coordinate space
$$
[X, Y]_{\cal R}=[{\cal R}(X), Y]+[X, {\cal R}(Y)]
$$
Now let us consider the induced symmetry transformation on
the functional space by gauge symmetry (6), which is in the
coordinate space.
$$
f_X(Q)\longrightarrow
G^{-1}f_X(Q)G\equiv f_X(G^{-1}QG)\in{\cal F}
$$
In particular, set $g=Q^k_{\alpha}$, and introduce the quantities
$$
H_k\equiv {1\over k}Tr(Q^k), \qquad\qquad  \forall k>0
$$
then
$$
\delta f_X(Q)=\epsilon X([Q, g])=\epsilon {1\over2}Q([
dH_{k+1}, X]_{\cal R})
$$
which can be considered as the transformation generated by
function $H_{k+1}$ through some Poisson bracket,
which is the first Poisson bracket
$$
\{f_X, f_Y\}_1(Q)={1\over2}Q([X, Y]_{\cal R})
\eqno\hbox{(22)}
$$
with respect to this Poisson bracket, $H_k$'s are
the conserved quantities.
Choosing one of them as Hamiltonian we recover the
KdV--Heirachioies.
Obviously, this Poisson bracket is antisymmetric and satisfies
the Jacobi identity.

In order to construct the second Poisson bracket, once again we use
 compatibility condition (10)(since it is valid for any integrable
system). Finally the straightforward computation shows that the
second Poisson bracket is
$$
\{ f_X, f_Y\}_2(Q)
={1\over2}<
QX{\cal R}(QY)-{\cal R}(YQ)XQ>
\eqno\hbox{(23)}
$$
Now let  us consider once again the n=2 case
$$
Q_{ij}=\sqrt{R_j}\delta_{j,i+1}+S_j\delta_{ij}+
l_j\delta_{j,i-1}+a_j\delta_{j,i-2}
$$
where
$$
\sqrt{R_j}l_j=L_j,\qquad\qquad \sqrt{R_jR_{j-1}}a_j=A_j
$$
then the Poisson bracket (23) gives the Poisson algebra
which contains (16) as a sub--algebra and  the following
ones
$$
\eqalignno{
\{S_i, R_j\}_2&=S_iR_j(\delta_{i,j-1}-\delta_{ij}),&(24a)\cr
\{L_i, R_j\}_2&=L_iR_j(\delta_{i,j-1}-\delta_{i,j+1}),&(24b)\cr
\{A_i, R_j\}_2&=A_iR_j(\delta_{i,j-1}-\delta_{i,j+2}),&(24c)\cr
\{R_i, R_j\}_2&=0,&(24d)\cr}
$$

There are a few unclear problems.
At first, it is interesting to investigate
the relation between
the continuum version of the above  Poisson brackets
and the Gelfand--Dickii Poisson brackets.
Secondly
these non--symmetric
lattices probably give a realization of discrete $W_n$--algebras.
One may also ask if there exist
some possible reduction of this model, which leaves only
one degree of freedom in $Q_{[-n,1]}, \forall n\geq2$. This
reduction might
be important in multi--matrix models[8].

\centerline{\bf Acknowledgements}

I would like to thank Prof. L.Bonora for valuable suggestions
and stimulating discussions.
I also wish to thank Prof.M.Martellini
for discussions and comments.

\vfill
\eject
\centerline{\bf References}
\vskip 1.0cm

\item{[1]} E.J.Martinec, ``On the Origin of Integrability of
Matrix Models", Chicago preprint, EFI--90--67

L.Alvarez--Gaume, ``The Integrability in Random Matrix Models",
CERN--TH.6123/91

\item{[2]} A.Gerasimov, A.Marshakov, A.Mironov, and
A.Orlov, ``Matrix Models of Two Dimensional Gravity and
Toda Theory", ITEP preprint (1990)

\item{[3]} L.Bonora, M.Martellini and C.S.Xiong, SISSA preprint
EP--107/91

\item{[4]} L.D.Faddeev, L.A.Takhtadzyan, ``Hamiltonian
Methods in the Theory of Solitons", Springer(1987)

\item{[5]} O.Babelon and C.Viallet, Lecture Notes in SISSA (1989)

\item{[6]} O.Babelon and L.Bonora, Phys.Lett.{\bf 253B}(1991)365

\item{[7]} J.Avan, Brown preprint BROWN--HET--812

\item{[8]} C.S.Xiong, SISSA preprint

\bye